\documentclass[twocolumn,showpacs,amsmath,amssymb,aps,prl,floatfix,superscriptaddress,10pt]{revtex4-2}

\usepackage[latin1]{inputenc}
\usepackage[american]{babel}
\usepackage[T1]{fontenc}
\usepackage{amsmath}
\usepackage{upgreek}
\usepackage{mhchem}
\usepackage[dvipsnames]{xcolor}
\usepackage{graphicx}
\usepackage{multirow}
\usepackage{physics}
\usepackage{changes}
\usepackage[caption=false]{subfig}

\usepackage{tikz}
\usetikzlibrary{decorations.pathmorphing}
\usetikzlibrary{shapes}
\usetikzlibrary{shapes.geometric}
\usepackage{tikz-feynman}

\begin{document}

\title{Evidence against nuclear polarization as source of fine-structure anomalies \\in muonic atoms}

\date{\today}

\author{Igor~A.~Valuev}
\email[Email: ]{igor.valuev@mpi-hd.mpg.de} 
\affiliation{Max-Planck-Institut f\"{u}r Kernphysik, Saupfercheckweg 1, 69117 Heidelberg, Germany}

\author{Gianluca~Col\`{o}}
\affiliation{Dipartimento di Fisica, Universit\`{a} degli Studi di Milano, via Celoria 16, I-20133 Milano, Italy}
\affiliation{INFN, Sezione di Milano, via Celoria 16, I-20133 Milano, Italy}

\author{Xavier~Roca-Maza}
\affiliation{Dipartimento di Fisica, Universit\`{a} degli Studi di Milano, via Celoria 16, I-20133 Milano, Italy}
\affiliation{INFN, Sezione di Milano, via Celoria 16, I-20133 Milano, Italy}

\author{Christoph~H.~Keitel}
\affiliation{Max-Planck-Institut f\"{u}r Kernphysik, Saupfercheckweg 1, 69117 Heidelberg, Germany}

\author{Natalia~S.~Oreshkina}
\email[Email: ]{natalia.oreshkina@mpi-hd.mpg.de} 
\affiliation{Max-Planck-Institut f\"{u}r Kernphysik, Saupfercheckweg 1, 69117 Heidelberg, Germany}

\begin{abstract}
A long-standing problem of fine-structure anomalies in muonic atoms is revisited by considering the $\Delta 2p$ splitting in muonic \ce{^{90}Zr}, \ce{^{120}Sn} and \ce{^{208}Pb} and the $\Delta 3p$ splitting in muonic \ce{^{208}Pb}.
State-of-the-art techniques from both nuclear and atomic physics are brought together in order to perform the most comprehensive to date calculations of nuclear-polarization energy shifts. 
Barring the more subtle case of $\mu$-\ce{^{208}Pb}, the results suggest that the dominant calculation uncertainty is much smaller than the persisting discrepancies between theory and experiment. 
We conclude that the resolution to the anomalies is likely to be rooted in refined QED corrections or even some other previously unaccounted-for contributions. 
\end{abstract}

\maketitle

\textit{Introduction.}---For more than 40 years there has been a perplexing discrepancy between theory and experiment with regard to the fine structure in muonic atoms~\cite{Yamazaki_muPb_1979, Bergem_muPb_1988, Phan_muZr_1985, Piller_muSn_1990}.
Due to the fact that $m_{\mu} \approx 207 m_e$, the Bohr radius of muonic orbitals is 207 times smaller than in ordinary electronic hydrogenlike atoms, which renders muon energy levels highly sensitive to nuclear structure~\cite{muon_data, proton_size, quadrupole_rare_earth, quadrupole_Re}.
In this respect, the most challenging effect to describe is the intricate interplay between muonic and internal nuclear degrees of freedom, which is known as nuclear polarization~(NP).
This phenomenon leads to shifts $\Delta E^{\mathrm{NP}}$ of muon energy levels, which can be observed in high-precision x\nobreakdash-ray measurements of muonic transitions.
Under the assumption that all other effects have been taken into account, the remaining difference between theory and experiment is typically ascribed to the NP correction.
However, in some cases, the NP energy shifts extracted in this way turned out to be in striking disagreement with theoretical predictions. 
For instance, the experiments suggest that $|\Delta E_{2p_{3/2}}^{\mathrm{NP}}| > |\Delta E_{2p_{1/2}}^{\mathrm{NP}}|$ for muonic~\ce{^{208}Pb}~\cite{Yamazaki_muPb_1979, Bergem_muPb_1988}, \ce{^{90}Zr}~\cite{Phan_muZr_1985} and \ce{^{112-124}Sn}~\cite{Piller_muSn_1990}.
At first glance, these results seem to be counterintuitive by a simple argument that the $2p_{1/2}$ orbital is closer to a nucleus and, thus, should be affected stronger by nuclear dynamics. 
In addition, a strong anomaly of the same kind has also been observed for the $\Delta 3p$ splitting in $\mu$\nobreakdash-\ce{^{208}Pb}~\cite{Bergem_muPb_1988}.

The most notable theoretical efforts to explain these anomalies were performed in Refs.~\cite{Tanaka_1994, Haga_mu_2002, Haga_mu_full_rel_2005, Haga_mu_2007}, where, unlike previous attempts, the transverse part of the electromagnetic muon-nucleus interaction was taken into account.
While the longitudinal, or Coulomb, part always leads to $|\Delta E_{2p_{3/2}}^{\mathrm{NP}}| < |\Delta E_{2p_{1/2}}^{\mathrm{NP}}|$ as expected, the transverse part was shown to give rise to an additional NP contribution with the opposite muon-spin dependency~\cite{Tanaka_1994}. 
According to Ref.~\cite{Haga_mu_2002}, the transverse interaction accounted for about half and one-fourth of the $\Delta 2p$ and $\Delta 3p$ anomalies in $\mu$\nobreakdash-\ce{^{208}Pb}, respectively. 
Nevertheless, significant portions of the discrepancies persisted, with $|\Delta E_{2p_{1/2}}^{\mathrm{NP}}|$ still being slightly larger than $|\Delta E_{2p_{3/2}}^{\mathrm{NP}}|$.
A glimpse of a possible resolution to the~$\Delta 2p$~anomaly in $\mu$\nobreakdash-\ce{^{208}Pb} was later provided in Ref.~\cite{Haga_mu_full_rel_2005} by treating the nucleus in the relativistic mean-field approximation.
However, the authors themselves stressed the large uncertainties associated with the nuclear spectrum obtained in this way, and explaining the~$\Delta 3p$~splitting still remained a challenge.
In another attempt the effect of enhancing the energy-weighted sum rule~(EWSR) with respect to its classical value was considered for both muonic \ce{^{208}Pb} and \ce{^{90}Zr}~\cite{Haga_mu_2007}. 
Once again, the experimental data could not be reproduced reasonably well, and the anomalies continued to be unresolved.

In this letter, we present a qualitative step forward in theoretical description of the NP effect by combining state-of-the-art tools in order to take into account both muonic and nuclear spectra in the most complete and precise to date manner.
The full electromagnetic muon-nucleus interaction is included within a field-theoretical framework.
Most importantly, nuclear model dependence is analyzed extensively leading to strong indications of NP not being responsible for the fine-structure anomalies in muonic atoms. 

\begin{figure}[!tbp]
\centering
\includegraphics[width=8.25cm]{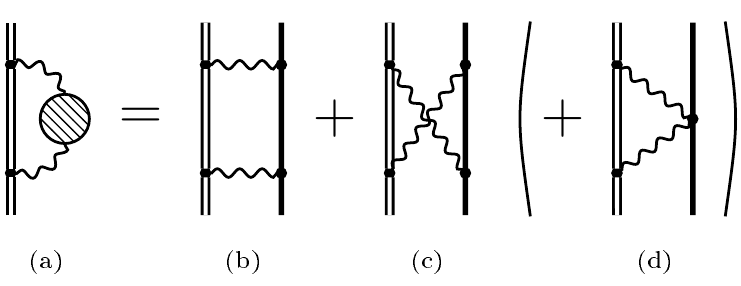}
\caption{Leading-order nuclear-polarization (NP) effect: (a)~effective self-energy Goldstone diagram with a dressed photon propagator; (b)~ladder, (c)~cross and (d)~seagull Feynman diagrams. A bound muon is denoted by a double line, while a nucleus is denoted by a single solid line. The shaded blob represents the NP insertion.}%
\label{fig:diagrams}
\end{figure}

\textit{Computational method.}---In the field-theoretical approach the NP effect can be described with an effective self-energy Goldstone diagram shown on Fig.~\ref{fig:diagrams}~(a).
The photon propagator $D_{\mu \nu}$ is modified by the so-called NP insertion, which is indicated as a shaded blob and can be expressed as~\cite{Plunien_1989}
\begin{align}
& \widetilde{D}_{\mu \nu}(x,x^{\prime}) = D_{\mu \nu}(x-x^{\prime}) \\
& + \int d^4 x_1 d^4 x_2 D_{\mu \xi}(x-x_1) \Pi ^{\xi \zeta}(x_1,x_2) D_{\zeta \nu}(x_2-x^{\prime}) \notag,
\end{align}
with the nuclear-polarization tensor
\begin{equation}
i\Pi^{\xi \zeta}(x_1,x_2) = \mel{\mathcal{O}}{\mathrm{T}[J^{\xi}_{N}(x_1) J^{\zeta}_{N}(x_2)]}{\mathcal{O}},
\end{equation}
where $J^{\mu}_{N}$ denotes the nuclear transition four-current density operator, and the ``vacuum'' state $\ket{\mathcal{O}}$ corresponds to the nucleus being in its ground state.
Here and later, four-vectors are represented by regular typeface, whereas three-vectors are denoted by bold letters.
The units $\hbar = c = 1$ and $\alpha = e^2/4\pi$ are used throughout the letter.

The leading-order NP effect can then be equivalently described by the ladder and cross Feynman diagrams representing a two-photon exchange between a bound muon and a nucleus~\cite{Mohr_1998} (Fig.~\ref{fig:diagrams}~(b) and~(c)).
However, if non-commuting nuclear charge and current operators are employed, it can be shown that an additional contribution has to be included in order to ensure gauge invariance of the NP correction~\cite{Haga_e_2002, Haga_rel_RPA}.
This additional term can be represented by the so-called seagull diagram (Fig.~\ref{fig:diagrams}~(d)), and in the case of non-relativistic nuclear charge-current operators it formally corresponds to the substitution~\cite{Haga_e_2002}:
\begin{equation}
\Pi^{\xi \zeta}(x_1,x_2) \rightarrow \dfrac{\mel{I}{\rho_N(\boldsymbol{x}_1)}{I}}{m_p} \delta^{\xi \zeta} \delta^{(4)}(x_1 - x_2),
\end{equation}
where $\ket{I}$ stands for the nuclear ground state, $\rho_{N}$ is the nuclear charge density operator, $m_p$ is the proton mass, and $\delta^{\xi \zeta}$ is the Kronecker delta extended to four dimensions with $\delta^{00}=0$.

The corresponding contributions to the NP energy shift for a muonic reference state $\ket{i}$ due to each of these diagrams (L, X and SG stand for ladder, cross and seagull, respectively) can be expressed in the momentum representation as~\cite{Haga_e_2002}
\begin{widetext}
\begin{align}
\label{eq:L}
\Delta E^{\mathrm{L}}_{\mathrm{NP}} & = -i (4\pi \alpha)^2 \sum_{i^{\prime} I^{\prime}} \iint \dfrac{d \boldsymbol{q} \, d \boldsymbol{q^{\prime}}}{(2\pi)^6} \int \dfrac{d \omega}{2\pi} \dfrac{D_{\mu \xi}(\omega, \boldsymbol{q}) D_{\zeta \nu}(\omega, \boldsymbol{q^{\prime}}) \mel{i I}{j^{\mu}_{m}(-\boldsymbol{q}) J^{\xi}_{N}(\boldsymbol{q})}{i^{\prime} I^{\prime}} \mel{i^{\prime} I^{\prime}}{J^{\zeta}_{N}(-\boldsymbol{q^{\prime}}) j^{\nu}_{m}(\boldsymbol{q^{\prime}})}{i I}}{(\omega + \omega_{m} - iE_{i^{\prime}}\epsilon)(\omega - \omega_{N} + i\epsilon)}, \\
\label{eq:X}
\Delta E^{\mathrm{X}}_{\mathrm{NP}} & = +i (4\pi \alpha)^2 \sum_{i^{\prime} I^{\prime}} \iint \dfrac{d \boldsymbol{q} \, d \boldsymbol{q^{\prime}}}{(2\pi)^6} \int \dfrac{d \omega}{2\pi} \dfrac{D_{\mu \xi}(\omega, \boldsymbol{q}) D_{\zeta \nu}(\omega, \boldsymbol{q^{\prime}}) \mel{i I^{\prime}}{j^{\mu}_{m}(-\boldsymbol{q}) J^{\xi}_{N}(\boldsymbol{q})}{i^{\prime} I} \mel{i^{\prime} I}{J^{\zeta}_{N}(-\boldsymbol{q^{\prime}}) j^{\nu}_{m}(\boldsymbol{q^{\prime}})}{i I^{\prime}}}{(\omega + \omega_{m} - iE_{i^{\prime}}\epsilon)(\omega + \omega_{N} - i\epsilon)}, \\
\label{eq:SG}
\Delta E^{\mathrm{SG}}_{\mathrm{NP}} & = -i (4\pi \alpha)^2 \sum_{i^{\prime}} \iint \dfrac{d \boldsymbol{q} \, d \boldsymbol{q^{\prime}}}{(2\pi)^6} \int \dfrac{d \omega}{2\pi} \dfrac{D_{\mu \xi}(\omega, \boldsymbol{q}) \delta^{\xi \zeta} D_{\zeta \nu}(\omega, \boldsymbol{q^{\prime}}) \mel{i}{j^{\mu}_{m}(-\boldsymbol{q})}{i^{\prime}} \mel{i^{\prime}}{j^{\nu}_{m}(\boldsymbol{q^{\prime}})}{i}}{(\omega + \omega_{m} - iE_{i^{\prime}}\epsilon)} \dfrac{\mel{I}{\rho_N(\boldsymbol{q}-\boldsymbol{q^{\prime}})}{I}}{m_p},
\end{align}
\end{widetext}
where the limit $\epsilon \rightarrow 0^{+}$ is implied, the indices $i^{\prime}$~and~$I^{\prime}$ in the sums run over an entire muonic Dirac spectrum and a complete set of nuclear excitations, respectively, $j^{\mu}_{m}$ is the Dirac four-current operator of the muon, $\omega_{m} = E_{i^{\prime}}-E_{i}$ and $\omega_{N} = E_{I^{\prime}}-E_{I}$. 
The specific formulas in the Feynman and Coulomb gauges are presented in Ref.~\cite{Haga_e_2002} (see the Supplemental Material for comments~\cite{supplement, Poincare_Bertrand}), and the expressions for the reduced matrix elements of both muonic and nuclear charge-current operators can be found in Ref.~\cite{Tanaka_1994}.

Taking into account a complete muonic Dirac spectrum poses a challenge since it includes an infinite set of bound states as well as positive- and negative-energy continua. 
Thus, direct calculations are difficult to implement with high accuracy, as they inevitably involve estimations of remainders of the sum over the bound states and the integrals over the continua.
In this work, we deal with this challenge by confining the system to a spherical cavity and employing finite basis-set expansions of the muon wave function in terms of B-splines~\cite{B_splines} within the dual-kinetic-balance approach~\cite{DKB}.
In this way, the continuous part of the spectrum becomes discrete, and the computation is reduced to finite sums with no remainders to evaluate.
The convergence of the results is readily controlled by varying the size of the cavity and the number of B-splines used.
The Dirac equation is solved in a potential of a nucleus with a finite charge distribution.
Similar to Ref.~\cite{Valuev_FNS}, we found that it is sufficient to use the simple Fermi charge distribution $\rho_{F}(r) = N \{ 1 + \mathrm{exp} [(r-c)/a] \}^{-1}$ with the standard value of the diffuseness parameter $a = 2.3/(4 \, \mathrm{ln}(3)) \, \mathrm{fm}$ and adjust the half-density radius $c$ such that a tabulated value of the root-mean-square nuclear radius~\cite{Nucl_rad} is~reproduced.

Computing a nuclear spectrum is yet more challenging since in the case of heavy nuclei an \textit{ab initio} description is not even feasible.
However, sophisticated particle-hole theories have proven to be very successful at describing the rich variety of nuclear excitations~\cite{2016_Nakatsukasa, 2007_Paar, 2018_RocaMaza}.
In our calculations we first carry out the Hartree-Fock computations of single-nucleon wave functions where the interactions between the nucleons are described by the Skyrme force~\cite{2003_Bender}.
Then we employ the random-phase approximation (RPA) with a full self-consistency between the Hartree-Fock mean field and the RPA excitations~\cite{skyrme_rpa}.
Non-relativistic charge-current operators~\cite{Tanaka_1994} are used for calculating the nuclear matrix elements in Eqs.~\eqref{eq:L}-\eqref{eq:SG} for the $0^{+}$, $1^{-}$, $2^{+}$, $3^{-}$, $4^{+}$, $5^{-}$ and $1^{+}$ excitation modes.
The~cutoff energy of the unoccupied single-particle states in the RPA model space is chosen to be 60~MeV, which corresponds, for example, to around 1500 RPA excitations in the case of the $3^{-}$ mode in \ce{^{208}Pb}.
The completeness of the obtained spectra is numerically controlled using the double-commutator EWSR, which is fulfilled at the level of at least 99.8\% in most cases. 
Thus, the present RPA description represents a significant improvement over the ones used previously in Refs.~\cite{Rinker_RPA, Tanaka_1994, Haga_mu_2002}.
Finally, parallel computing on a cluster is employed to achieve high precision in such combined muon-nuclear calculations.

\textit{Results and discussion.}---In Table~\ref{table:1} we present our results for NP corrections (absolute values $|\Delta E^{\mathrm{NP}}|=-\Delta E^{\mathrm{NP}}$) to the ground state $1s_{1/2}$ as well as the excited states $2p_{1/2}$ and $2p_{3/2}$ in muonic \ce{^{90}Zr}, \ce{^{120}Sn} and \ce{^{208}Pb}.
In the case of $\mu$-\ce{^{208}Pb} the states $3p_{1/2}$ and $3p_{3/2}$ are also considered. 
The quantities of main interest are the corresponding NP contributions to the fine-structure splittings $\Delta 2p^{\mathrm{NP}} = |\Delta E_{2p_{1/2}}^{\mathrm{NP}}| - |\Delta E_{2p_{3/2}}^{\mathrm{NP}}|$ and $\Delta 3p^{\mathrm{NP}} = |\Delta E_{3p_{1/2}}^{\mathrm{NP}}| - |\Delta E_{3p_{3/2}}^{\mathrm{NP}}|$.
Our calculations in the Feynman and Coulomb gauges agree within 0.1\nobreakdash--0.3\% demonstrating an excellent fulfillment of gauge invariance.
Table~\ref{table:1} contains total NP~corrections in the Feynman gauge, while separate contributions from each type of nuclear excitations are listed in the Supplemental Material~\cite{supplement}.

The main limitation of any NP calculation comes from the fact that nuclear transition charge and current densities are not known from first principles.
As a consequence, an effective nuclear model has to be applied, and the NP correction inevitably becomes model dependent.
In this work, we analyze this model dependence extensively by performing the computations for 9 different Skyrme parametrizations, namely, KDE0, SKX, SLy5, BSk14, SAMi, NRAPR, SkP, SkM* and SGII, covering a wide range in the parameter space~\cite{KDE0, SKX, SLy5, BSk14, SAMi, NRAPR, SkP, SkM, SGII}.

\begin{figure}[!htbp]
\centering
\includegraphics[width=8.5cm]{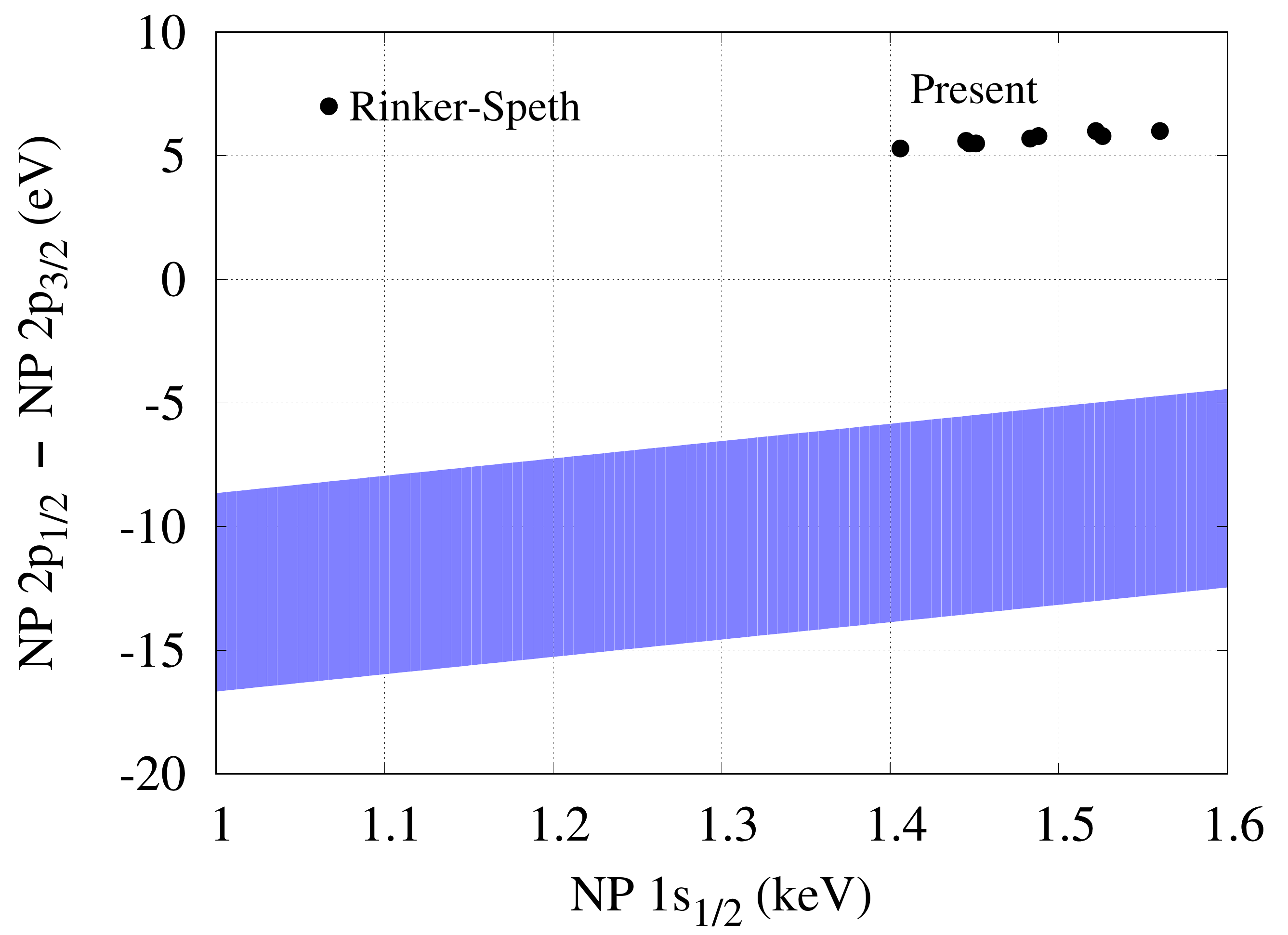}
\caption{Theoretical values of the nuclear-polarization~(NP) corrections for $\mu$-\ce{^{90}Zr} in relation to the experimentally allowed range for $\Delta 2p^{\mathrm{NP}}$ as a function of $|\Delta E_{1s_{1/2}}^{\mathrm{NP}}|$. The graph was adapted from Ref.~\cite{Phan_muZr_1985}.}%
\label{fig:zr}
\end{figure}

We start our analysis with perhaps the most prominent case of $\mu$-\ce{^{90}Zr}. 
To put the effect of nuclear model dependence into the context of the $\Delta 2p$ fine-structure anomaly, we show our results on Fig.~\ref{fig:zr} in relation to the experimentally allowed region~\cite{Phan_muZr_1985} for $|\Delta E_{1s_{1/2}}^{\mathrm{NP}}|$ and $\Delta 2p^{\mathrm{NP}}$.
Most notably, the results for different nuclear models are simply spread along a line almost parallel to the allowed region.
Thus, even though individual NP corrections can vary significantly depending on the Skyrme interaction, the distance between theory and experiment for $\Delta 2p^{\mathrm{NP}}$ remains practically constant. 

\begin{table*}[!htbp]
\caption{Nuclear-polarization (NP) corrections (absolute values $|\Delta E^{\mathrm{NP}}|=-\Delta E^{\mathrm{NP}}$, in~eV) to the states $1s_{1/2}$, $2p_{1/2}$ and $2p_{3/2}$ in muonic \ce{^{90}Zr}, \ce{^{120}Sn} and \ce{^{208}Pb}. In the case of~$\mu$-\ce{^{208}Pb} the states $3p_{1/2}$ and $3p_{3/2}$ are also considered. The quantities $\Delta 2p^{\mathrm{NP}} = |\Delta E_{2p_{1/2}}^{\mathrm{NP}}| - |\Delta E_{2p_{3/2}}^{\mathrm{NP}}|$ and $\Delta 3p^{\mathrm{NP}} = |\Delta E_{3p_{1/2}}^{\mathrm{NP}}| - |\Delta E_{3p_{3/2}}^{\mathrm{NP}}|$ are the corresponding NP contributions to the fine-structure splittings. The Skyrme parametrizations are ordered in increasing values of the ground-state correction in~$\mu$-\ce{^{90}Zr}.}
\label{table:1}
{\renewcommand{\arraystretch}{1.1}
\renewcommand{\tabcolsep}{0.3838cm}
\begin{tabular}{llr@{}lr@{}lr@{}lr@{}lr@{}lr@{}lr@{}lr@{}lr@{}l}
\hline \hline
&& \multicolumn{2}{c}{KDE0} & \multicolumn{2}{c}{SKX} & \multicolumn{2}{c}{SLy5} & \multicolumn{2}{c}{BSk14} & \multicolumn{2}{c}{SAMi} & \multicolumn{2}{c}{NRAPR} & \multicolumn{2}{c}{SkP} & \multicolumn{2}{c}{SkM*} & \multicolumn{2}{c}{SGII} \\
\hline                                                                                        
$\mu$-\ce{^{90}Zr} & $1s_{1/2}$ & \phantom{.}140&6 & 144&5 & 144&7 & \phantom{)}145&1 & \phantom{.}148&3 & \phantom{0.}148&8 & 152&2 & 152&6 & 156&0 \\
&                    $2p_{1/2}$ & 65.&9 & 70.&3 & 69.&5 & 70.&0 & 72.&5 & 71.&7 & 73.&9 & 74.&4 & 75.&7 \\
&                    $2p_{3/2}$ & 60.&6 & 64.&7 & 64.&0 & 64.&5 & 66.&8 & 65.&9 & 67.&9 & 68.&6 & 69.&7 \\
&     $\Delta 2p^{\mathrm{NP}}$ &  5.&3 &  5.&6 &  5.&5 &  5.&5 &  5.&7 &  5.&8 &  6.&0 &  5.&8 &  6.&0 \\[1mm]
\hline                                                                                
$\mu$-\ce{^{120}Sn}& $1s_{1/2}$ & 256&4 & 251&0 & 248&1 & 242&5 & 253&0 & 253&1 & 257&0 & 256&7 & 274&4 \\
&                    $2p_{1/2}$ &  24&7 &  24&8 &  23&6 &  23&1 &  24&6 &  24&5 &  24&7 &  24&7 &  26&9 \\
&                    $2p_{3/2}$ &  22&8 &  22&9 &  21&8 &  21&4 &  22&8 &  22&6 &  22&7 &  22&8 &  24&8 \\
&     $\Delta 2p^{\mathrm{NP}}$ & 19.&9 & 19.&6 & 18.&0 & 17.&0 & 18.&7 & 18.&7 & 19.&2 & 18.&9 & 21.&1 \\[1mm]
\hline                                                                                
$\mu$-\ce{^{208}Pb}& $1s_{1/2}$ & 546&3 & 543&2 & 555&7 & 558&8 & 572&7 & 588&9 & 581&5 & 590&5 & 603&5 \\
&                    $2p_{1/2}$ & 178&1 & 185&0 & 183&4 & 190&0 & 193&7 & 199&7 & 195&5 & 200&5 & 204&4 \\
&                    $2p_{3/2}$ & 172&5 & 179&8 & 177&6 & 185&2 & 187&7 & 193&6 & 188&6 & 194&2 & 198&1 \\
&                    $3p_{1/2}$ &  52&9 &  57&6 &  55&6 &  56&6 &  61&6 &  54&0 &  62&8 &  61&4 &  62&7 \\
&                    $3p_{3/2}$ &  55&9 &  61&2 &  58&9 &  60&2 &  64&8 &  57&6 &  67&2 &  64&5 &  66&4 \\
&     $\Delta 2p^{\mathrm{NP}}$ & 56.&0 & 51.&8 & 57.&5 & 48.&1 & 59.&1 & 60.&5 & 69.&3 & 63.&3 & 62.&7 \\
&     $\Delta 3p^{\mathrm{NP}}$ &-29.&5 &-35.&9 &-33.&4 &-36.&1 &-31.&9 &-35.&8 &-44.&1 &-30.&3 &-37.&3 \\
\hline \hline
\end{tabular}}
\end{table*}

\begin{figure*}[!htbp]%
\centering
\subfloat[\centering]{{\includegraphics[width=8.5cm]{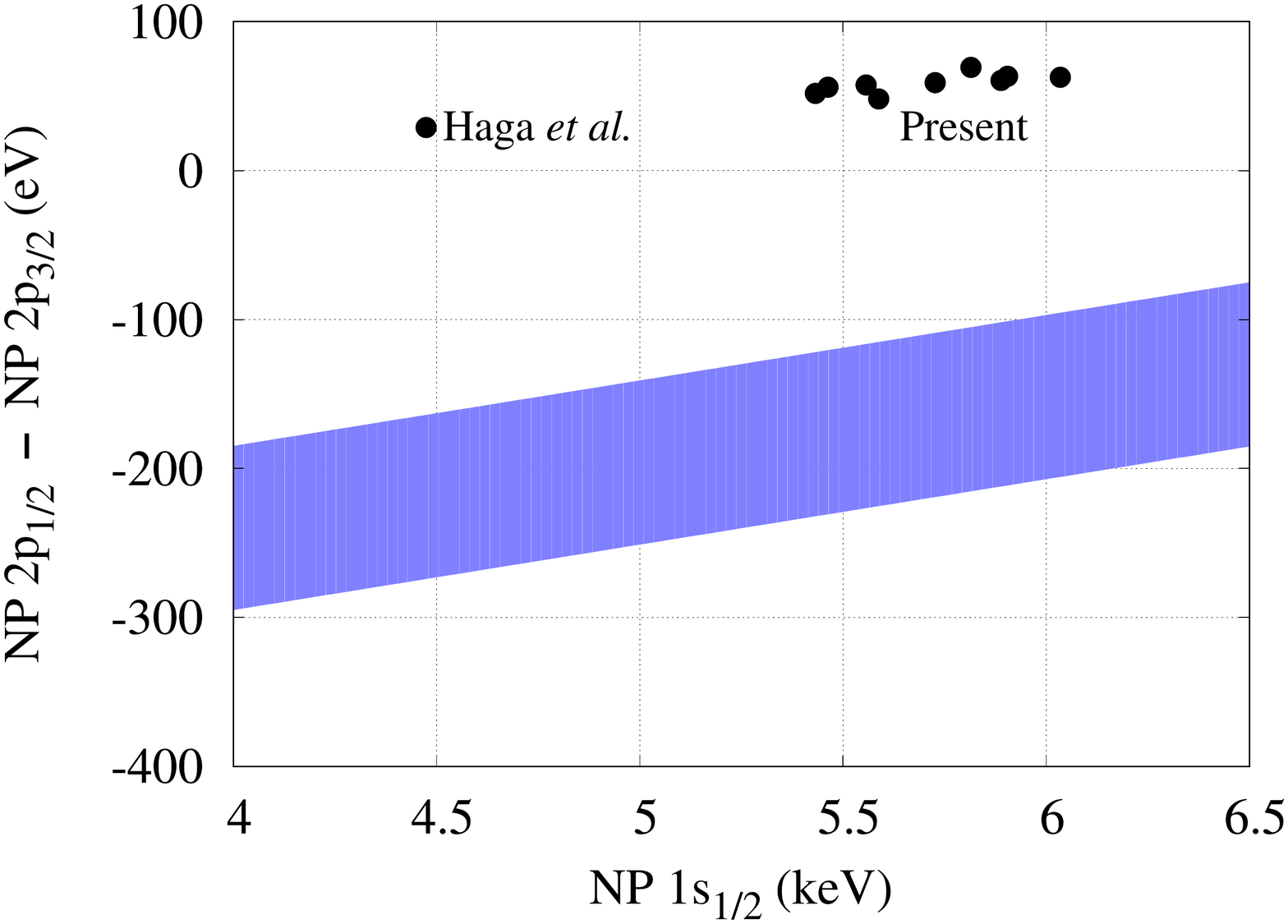} }}%
\qquad
\subfloat[\centering]{{\includegraphics[width=8.5cm]{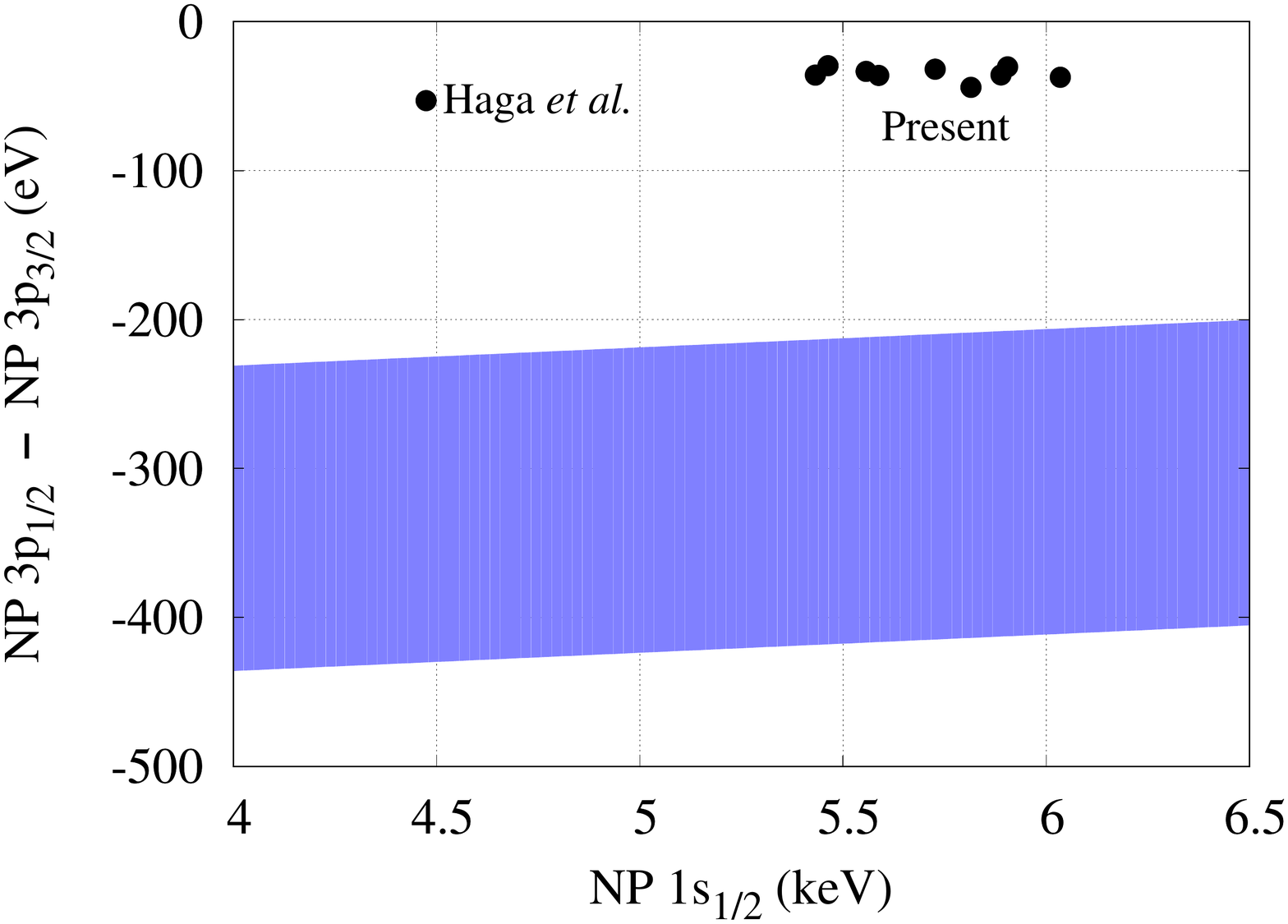} }}%
\caption{Theoretical values of the nuclear-polarization~(NP) corrections for $\mu$-\ce{^{208}Pb} in relation to the experimentally allowed ranges for $\Delta 2p^{\mathrm{NP}}$~(a) and $\Delta 3p^{\mathrm{NP}}$~(b) as functions of $|\Delta E_{1s_{1/2}}^{\mathrm{NP}}|$. The graphs were adapted from Ref.~\cite{Haga_mu_2002}.}%
\label{fig:pb}%
\end{figure*}

As for tin isotopes, the authors of Ref.~\cite{Piller_muSn_1990} do not provide experimentally allowed ranges for $\Delta 2p^{\mathrm{NP}}$. 
Nevertheless, according to their analysis, the theoretical values of the $\Delta 2p$ fine-structure splittings are consistently too high by about 150~eV, and furthermore, it is necessary to have $\Delta 2p^{\mathrm{NP}} < 0$ in order to obtain a better agreement with experiment.
However, the authors estimate $\Delta 2p^{\mathrm{NP}}$ as 29~eV and 28~eV for muonic \ce{^{112}Sn} and \ce{^{124}Sn}, respectively.
Our results for $\mu$\nobreakdash-\ce{^{120}Sn} in Table~\ref{table:1} demonstrate again that the nuclear model uncertainty does not offer an explanation for the anomalies, with $\Delta 2p^{\mathrm{NP}}$ being persistently positive and around 20~eV for all the Skyrme parametrizations used.

In the case of $\mu$-\ce{^{208}Pb} the situation is more subtle since, in principle, some $1^{-}$ nuclear excitations in the regions 5.5--6.5~MeV and 8--9~MeV~\cite{Nucl_data_208} may come close in energy to the $2p \rightarrow 1s$ and $3p \rightarrow 1s$ muonic transitions, respectively, resulting in muon-nuclear resonances.
This phenomenon was first noticed in Ref.~\cite{Shakin_resonances} for the $3d \rightarrow 2p$ muonic transitions and the low-lying $3^{-}$ nuclear state at 2.615~MeV.
With regard to $1^{-}$ resonant levels, the effect is even stronger due to the long range of the dipole NP potential.
As it was discussed in Ref.~\cite{Rinker_RPA}, $1^{-}$ muon-nuclear resonances can be significant even when the associated energy denominators in a second-order perturbed calculation are hundreds of~keV.
The net effect is highly sensitive not only to the exact relative positions of the muonic and nuclear levels involved but also to the shapes of the corresponding nuclear transition charge and current densities~\cite{Haga_mu_2007}.

In our calculated spectra for~\ce{^{208}Pb} we encounter a number of $1^{-}$ excitations in both aforementioned regions. 
Although RPA is an excellent tool for describing integral properties of a nuclear spectrum as a whole, the accuracy for individual energy levels is by no means high enough to reliably predict such resonant phenomena.
Therefore, similar to Ref.~\cite{Rinker_RPA}, we simply eliminate any accidental muon-nuclear resonances by discarding $1^{-}$ RPA excitations that come closer than 0.3 MeV to the $2p \rightarrow 1s$ or $3p \rightarrow 1s$ muonic transitions. 
It is worth noting, however, that this does not significantly affect the overall completeness of the spectra, since the total contributions of the discarded RPA states to the EWSR are always less~than~1\%.
Fig.~\ref{fig:pb} shows the resulting NP correlations between $|\Delta E_{1s_{1/2}}^{\mathrm{NP}}|$ and both $\Delta 2p^{\mathrm{NP}}$~(a) and $\Delta 3p^{\mathrm{NP}}$~(b) in relation to the experimentally allowed regions~\cite{Bergem_muPb_1988}.
It can be seen that, in the absence of muon-nuclear resonances, the model uncertainty is once again much smaller than the gap between theory and experiment.
We emphasize that due to the extremely high intrinsic uncertainties associated with muon-nuclear resonances, they should be regarded as a measure of last resort in explaining the fine-structure anomalies in~$\mu$-\ce{^{208}Pb}, and their treatment goes beyond the scope of this letter. 

\textit{Conclusions and outlook.}---In the quest to explain various persisting discrepancies between theory and experiment, we have performed the most complete to date calculations of the NP effect in muonic \ce{^{90}Zr}, \ce{^{120}Sn} and \ce{^{208}Pb}. 
Utilizing state-of-the-art techniques and leveraging modern computational power allows us to take into account the entire muonic and nuclear spectra in a controlled manner and with an improved precision.

We have found that the dominant nuclear model uncertainty is of surprisingly minor importance in the context of the fine-structure anomalies in muonic atoms, leading to even more tension between theory and experiment.
One should bear in mind possible complications in the special case of $\mu$-\ce{^{208}Pb} due to potential muon-nuclear resonances; therefore, we suggest that the less intricate cases of muonic~\ce{^{90}Zr} and \ce{^{112-124}Sn} should be tackled first.
The non-relativistic nuclear treatment in our calculations is justified by the excellent agreement between the non-relativistic seagull term and antinucleon NP contributions in light muonic atoms~\cite{Haga_rel_RPA}.
In addition, there is a general consistency between relativistic and non-relativistic approaches for a variety of nuclear phenomena~\cite{2007_Paar, 2018_RocaMaza, 2003_Bender}. 
However, in the special case of NP, a~possible non-negligible role of relativistic nuclear effects in heavy systems may still deserve further investigation, as proposed in Refs.~\cite{Haga_mu_full_rel_2005, Haga_rel_RPA}.

For the most part, we deem the NP effect unlikely to be responsible for the anomalies, implying that the solution is presumably rooted in refined QED calculations.
In particular, the self-energy correction in muonic atoms, despite being comparable to the NP shifts~\cite{Haga_mu_2007}, has only been estimated using rather simple prescriptions~\cite{Borie_Rinker}.
Therefore, a rigorous treatment of this effect developed in the field of highly-charged ions (see, e.g., Refs.~\cite{Indelicato_self_energy, Snyderman_self_energy, Yerokhin_self_energy}) could shed some light on the anomalies.
Lastly, some other exotic effects, such as the anomalous spin-dependent interaction mentioned in Ref.~\cite{Rinker_exotic}, might also play a role in explaining the discrepancies, although it is far less likely.
In summary, we conclude that more attention to other effects beyond NP is required in order to finally resolve this tantalizing and long-standing puzzle.

\textit{Acknowledgements.}---This article comprises parts of the PhD thesis work of I.A.V. to be submitted to the Heidelberg University, Germany.
The authors thank N.~Minkov, H.~Cakir, V.~A.~Yerokhin and Z.~Harman for helpful discussions.

\end{document}


\title{Supplemental Material for\\
``Evidence against nuclear polarization as source of fine-structure anomalies\\ in muonic atoms''}

\author{Igor~A.~Valuev}
\email[Email: ]{igor.valuev@mpi-hd.mpg.de} 
\affiliation{Max-Planck-Institut f\"{u}r Kernphysik, Saupfercheckweg 1, 69117 Heidelberg, Germany}

\author{Gianluca~Col\`{o}}
\affiliation{Dipartimento di Fisica, Universit\`{a} degli Studi di Milano, via Celoria 16, I-20133 Milano, Italy}
\affiliation{INFN, Sezione di Milano, via Celoria 16, I-20133 Milano, Italy}

\author{Xavier~Roca-Maza}
\affiliation{Dipartimento di Fisica, Universit\`{a} degli Studi di Milano, via Celoria 16, I-20133 Milano, Italy}
\affiliation{INFN, Sezione di Milano, via Celoria 16, I-20133 Milano, Italy}

\author{Christoph~H.~Keitel}
\affiliation{Max-Planck-Institut f\"{u}r Kernphysik, Saupfercheckweg 1, 69117 Heidelberg, Germany}

\author{Natalia~S.~Oreshkina}
\email[Email: ]{natalia.oreshkina@mpi-hd.mpg.de} 
\affiliation{Max-Planck-Institut f\"{u}r Kernphysik, Saupfercheckweg 1, 69117 Heidelberg, Germany}

\maketitle

\section{1 Comments on the formulas presented in Ref.~\cite{Haga_e_2002}}

\subsection{1.1 The Coulomb gauge}

1) Please note a minor misprint in the formula~(23) in Ref.~\cite{Haga_e_2002} for the $D^{C}_{00}$ component of the photon propagator. The correct expression is
\begin{equation}
D^{C}_{00}(\omega, \boldsymbol{q}) = \dfrac{1}{\boldsymbol{q}^{2} + i\epsilon}, 
\end{equation}
with a three-vector $\boldsymbol{q}$ in the denominator instead of a four-vector $q$.
\vskip 2mm

2) There is also a small error in the formula~(30). It should read:
\begin{equation}
\mathcal{W}^{C}_{T}(q) = -\sum_{\lambda} \left[ \dfrac{\omega_{e} \omega_{N}}{q^2} \mel{i^{\prime}}{|m_{\lambda}(q)|}{i} \mel{I^{\prime}}{|M_{\lambda}(q)|}{I} \ + \sum_{L=\lambda-1}^{\lambda+1} \mel{i^{\prime}}{|t_{\lambda L}(q)|}{i} \mel{I^{\prime}}{|T_{\lambda L}(q)|}{I} \right],
\end{equation}
such that there is no additional sum over $\lambda$ inside the square brackets.

\subsection{1.2 Corrections to excited muon states}

We would like to point out a subtlety with regard to calculating nuclear-polarization energy shifts to excited muon states. In these cases one encounters integrals of the following form:
\begin{equation}
\Delta E = \int_{0}^{\infty} \int_{0}^{\infty} \frac{f(q,q^{\prime}) \, dq \, dq^{\prime}}{(q-a-i0)(q^{\prime}-b-i0)},
\end{equation}
where $a>0$ and $b>0$.
In evaluating such two-dimensional integrals, due to the Poincar\'{e}-Bertrand theorem, there is an additional term as compared to simply applying the Sokhotski-Plemelj formula twice~\cite{Poincare_Bertrand}:
\begin{align}
& \frac{1}{q-a-i0}\frac{1}{q^{\prime}-b-i0} = \notag \\
& \left[ \frac{\mathcal{P}}{q-a} + i \pi \delta(q-a) \right] \left[ \frac{\mathcal{P}}{q^{\prime}-b} + i \pi \delta(q^{\prime}-b) \right] \notag \\
& + \pi^2 \delta(q-a)\delta(q^{\prime}-b),
\end{align}
such that the products of the delta-functions cancel each other resulting in
\begin{align}
\mathrm{Re}(\Delta E) = \mathcal{P} \int_{0}^{\infty} \mathcal{P} \int_{0}^{\infty} \frac{f(q,q^{\prime}) \, dq \, dq^{\prime}}{(q-a)(q^{\prime}-b)},
\end{align}
where $\mathcal{P}$ denotes the Cauchy principal value.

\newpage

\section{2 Detailed numerical results on nuclear model dependence}

\subsection{2.1 $\boldsymbol{\mu}$-\ce{^{90}Zr}}

\begin{table*}[!hbtp]
\caption{Contributions from different nuclear excitation modes to the nuclear-polarization~(NP) corrections (absolute values $|\Delta E^{\mathrm{NP}}|=-\Delta E^{\mathrm{NP}}$, in~eV) to the states $1s_{1/2}$, $2p_{1/2}$ and $2p_{3/2}$ in muonic \ce{^{90}Zr}. The quantity $\Delta 2p^{\mathrm{NP}} = |\Delta E_{2p_{1/2}}^{\mathrm{NP}}| - |\Delta E_{2p_{3/2}}^{\mathrm{NP}}|$ is the corresponding NP contribution to the fine-structure splitting. The values are given for 9 different Skyrme parametrizations.}
{\renewcommand{\arraystretch}{1.35}
\renewcommand{\tabcolsep}{0.3636cm}
\begin{tabular}{lcr@{}lr@{}lr@{}lr@{}lr@{}lr@{}lr@{}lr@{}lr@{}l}
\hline \hline
           &          & \multicolumn{2}{c}{KDE0}  & \multicolumn{2}{c}{SKX}   & \multicolumn{2}{c}{SLy5}  & \multicolumn{2}{c}{BSk14} & \multicolumn{2}{c}{SAMi}  & \multicolumn{2}{c}{NRAPR} & \multicolumn{2}{c}{SkP}   & \multicolumn{2}{c}{SkM*}  & \multicolumn{2}{c}{SGII}  \\
\hline                                                                                        
$1s_{1/2}$ & $0^{+}$  & 233.&2 & 218.&9 & 243.&5 & 230.&1 & 239.&3 & \phantom{0)}245.&7 & 261.&4 & 253.&8 & 253.&2 \\
           & $1^{-}$  & 637.&0 & 646.&1 & 666.&2 & 658.&9 & 706.&5 & 692.&2 & 679.&4 & 708.&8 & 728.&2 \\
           & $2^{+}$  & 308.&5 & 345.&3 & 311.&4 & 322.&4 & 310.&3 & 315.&2 & 355.&2 & 329.&9 & 332.&0 \\
           & $3^{-}$  & 160.&0 & 162.&5 & 157.&5 & 171.&9 & 159.&6 & 164.&4 & 153.&2 & 163.&7 & 174.&1 \\
           & $4^{+}$  & 42.&4  & 46.&3  & 42.&9  & 43.&3  & 42.&8  & 43.&9  & 46.&8  & 44.&6  & 45.&8  \\
           & $5^{-}$  & 20.&1  & 21.&4  & 20.&5  & 20.&1  & 20.&3  & 20.&9  & 21.&6  & 20.&9  & 21.&7  \\
           & $1^{+}$  & 4.&4   & 4.&8   & 4.&5   & 4.&5   & 4.&6   & 5.&4   & 4.&9   & 4.&5   & 4.&6   \\
\hline                                                                               
           & Total    &1405.&6 &1445.&4 &1446.&6 &1451.&1 &1483.&3 &1487.&6 &1522.&4 &1526.&2 &1559.&5 \\[2mm]
\hline                                                                               
$2p_{1/2}$ & $0^{+}$  & 1.&8   & 1.&7   & 1.&9   & 1.&8   & 1.&9   & 1.&9   & 2.&0   & 2.&0   & 1.&9   \\
           & $1^{-}$  & 38.&6  & 39.&9  & 41.&6  & 40.&8  & 44.&7  & 43.&0  & 42.&6  & 45.&0  & 45.&9  \\
           & $2^{+}$  & 15.&6  & 18.&5  & 16.&1  & 16.&7  & 15.&9  & 16.&5  & 19.&5  & 17.&2  & 17.&0  \\
           & $3^{-}$  & 7.&6   & 7.&7   & 7.&6   & 8.&4   & 7.&6   & 7.&8   & 7.&2   & 7.&9   & 8.&4   \\
           & $4^{+}$  & 1.&5   & 1.&7   & 1.&6   & 1.&6   & 1.&6   & 1.&6   & 1.&7   & 1.&6   & 1.&7   \\
           & $5^{-}$  & 0.&7   & 0.&7   & 0.&7   & 0.&7   & 0.&7   & 0.&7   & 0.&8   & 0.&7   & 0.&7   \\
           & $1^{+}$  & 0.&1   & 0.&1   & 0.&1   & 0.&1   & 0.&1   & 0.&2   & 0.&1   & 0.&1   & 0.&1   \\
\hline                                                                               
           & Total    & 65.&9  & 70.&3  & 69.&5  & 70.&0  & 72.&5  & 71.&7  & 73.&9  & 74.&4  & 75.&7  \\[2mm]
\hline                                                                               
$2p_{3/2}$ & $0^{+}$  & 0.&8   & 0.&8   & 0.&9   & 0.&8   & 0.&9   & 0.&9   & 1.&0   & 0.&9   & 0.&9   \\
           & $1^{-}$  & 36.&6  & 37.&8  & 39.&5  & 38.&8  & 42.&4  & 40.&8  & 40.&3  & 42.&7  & 43.&6  \\
           & $2^{+}$  & 14.&5  & 17.&1  & 14.&9  & 15.&5  & 14.&7  & 15.&3  & 18.&1  & 15.&9  & 15.&8  \\
           & $3^{-}$  & 6.&7   & 6.&8   & 6.&6   & 7.&4   & 6.&7   & 6.&9   & 6.&4   & 6.&9   & 7.&3   \\
           & $4^{+}$  & 1.&3   & 1.&4   & 1.&3   & 1.&3   & 1.&3   & 1.&4   & 1.&5   & 1.&4   & 1.&4   \\
           & $5^{-}$  & 0.&6   & 0.&6   & 0.&6   & 0.&6   & 0.&6   & 0.&6   & 0.&6   & 0.&6   & 0.&6   \\
           & $1^{+}$  & 0.&1   & 0.&1   & 0.&1   & 0.&1   & 0.&1   & 0.&1   & 0.&1   & 0.&1   & 0.&1   \\
\hline                                                                               
           & Total    & 60.&6  & 64.&7  & 64.&0  & 64.&5  & 66.&8  & 65.&9  & 67.&9  & 68.&6  & 69.&7  \\[2mm]
\hline                                                                               
$\Delta 2p$&          & 5.&3   & 5.&6   & 5.&5   & 5.&5   & 5.&7   & 5.&8   & 6.&0   & 5.&8   & 6.&0   \\
\hline \hline
\end{tabular}}
\end{table*}

\newpage

\subsection{2.2 $\boldsymbol{\mu}$-\ce{^{120}Sn}}

\begin{table*}[!hbtp]
\caption{Contributions from different nuclear excitation modes to the nuclear-polarization~(NP) corrections (absolute values $|\Delta E^{\mathrm{NP}}|=-\Delta E^{\mathrm{NP}}$, in~eV) to the states $1s_{1/2}$, $2p_{1/2}$ and $2p_{3/2}$ in muonic \ce{^{120}Sn}. The quantity $\Delta 2p^{\mathrm{NP}} = |\Delta E_{2p_{1/2}}^{\mathrm{NP}}| - |\Delta E_{2p_{3/2}}^{\mathrm{NP}}|$ is the corresponding NP contribution to the fine-structure splitting. The values are given for 9 different Skyrme parametrizations.}
{\renewcommand{\arraystretch}{1.35}
\renewcommand{\tabcolsep}{0.3636cm}
\begin{tabular}{lcr@{}lr@{}lr@{}lr@{}lr@{}lr@{}lr@{}lr@{}lr@{}l}
\hline \hline
           &          & \multicolumn{2}{c}{KDE0}  & \multicolumn{2}{c}{SKX}   & \multicolumn{2}{c}{SLy5}  & \multicolumn{2}{c}{BSk14} & \multicolumn{2}{c}{SAMi}  & \multicolumn{2}{c}{NRAPR} & \multicolumn{2}{c}{SkP}   & \multicolumn{2}{c}{SkM*}  & \multicolumn{2}{c}{SGII}  \\
\hline                                                                                        
$1s_{1/2}$ & $0^{+}$  & 425.&5 & 389.&6 & 441.&6 & 422.&0 & 430.&4 & \phantom{0)}443.&6 & 470.&9 & 465.&5 & 462.&3 \\
           & $1^{-}$  & 1060.&4& 1066.&4& 1087.&7& 1112.&7& 1158.&2& 1130.&0& 1104.&2& 1177.&7& 1202.&1\\
           & $2^{+}$  & 538.&0 & 582.&0 & 541.&2 & 561.&9 & 546.&0 & 555.&1 & 601.&1 & 582.&0 & 585.&3 \\
           & $3^{-}$  & 423.&3 & 345.&4 & 291.&0 & 208.&7 & 276.&6 & 278.&7 & 266.&7 & 219.&0 & 364.&5 \\
           & $4^{+}$  & 76.&6  & 82.&7  & 77.&0  & 78.&8  & 77.&4  & 79.&1  & 82.&4  & 80.&8  & 83.&4  \\
           & $5^{-}$  & 33.&7  & 36.&8  & 35.&2  & 34.&2  & 34.&6  & 35.&5  & 37.&4  & 35.&4  & 38.&9  \\
           & $1^{+}$  & 6.&8   & 7.&5   & 6.&9   & 6.&9   & 7.&0   & 8.&8   & 7.&6   & 6.&9   & 7.&1   \\
\hline                                                                               
           & Total    &2564.&2 &2510.&4 &2480.&6 &2425.&3 &2530.&2 &2530.&8 &2570.&4 &2567.&3 &2743.&6 \\[2mm]
\hline                                                                               
$2p_{1/2}$ & $0^{+}$  & 6.&9   & 6.&4   & 7.&2   & 6.&9   & 7.&1   & 7.&2   & 7.&7   & 7.&6   & 7.&5   \\
           & $1^{-}$  & 121.&9 & 125.&7 & 127.&6 & 132.&1 & 139.&2 & 133.&5 & 130.&1 & 142.&3 & 144.&3 \\
           & $2^{+}$  & 59.&4  & 67.&4  & 60.&9  & 63.&1  & 61.&9  & 65.&4  & 71.&5  & 66.&6  & 66.&4  \\
           & $3^{-}$  & 50.&5  & 39.&1  & 31.&3  & 19.&7  & 29.&0  & 29.&1  & 27.&5  & 21.&0  & 41.&2  \\
           & $4^{+}$  & 5.&9   & 6.&5   & 6.&1   & 6.&2   & 6.&1   & 6.&3   & 6.&5   & 6.&4   & 6.&5   \\
           & $5^{-}$  & 2.&5   & 2.&7   & 2.&6   & 2.&5   & 2.&6   & 2.&6   & 2.&8   & 2.&6   & 2.&9   \\
           & $1^{+}$  & 0.&4   & 0.&5   & 0.&4   & 0.&4   & 0.&4   & 0.&7   & 0.&5   & 0.&5   & 0.&5   \\
\hline                                                                               
           & Total    & 247.&4 & 248.&2 & 236.&1 & 230.&8 & 246.&3 & 244.&8 & 246.&5 & 246.&9 & 269.&2 \\[2mm]
\hline                                                                               
$2p_{3/2}$ & $0^{+}$  & 3.&4   & 3.&2   & 3.&6   & 3.&4   & 3.&5   & 3.&6   & 3.&8   & 3.&8   & 3.&7   \\
           & $1^{-}$  & 115.&8 & 119.&0 & 121.&3 & 125.&7 & 132.&0 & 126.&4 & 123.&1 & 134.&9 & 136.&9 \\
           & $2^{+}$  & 55.&9  & 63.&4  & 57.&5  & 59.&4  & 58.&4  & 62.&1  & 67.&5  & 62.&7  & 62.&4  \\
           & $3^{-}$  & 44.&9  & 34.&8  & 27.&9  & 17.&5  & 25.&9  & 25.&9  & 24.&6  & 18.&7  & 36.&6  \\
           & $4^{+}$  & 5.&2   & 5.&6   & 5.&3   & 5.&3   & 5.&3   & 5.&5   & 5.&6   & 5.&5   & 5.&7   \\
           & $5^{-}$  & 2.&1   & 2.&2   & 2.&2   & 2.&1   & 2.&2   & 2.&2   & 2.&3   & 2.&2   & 2.&4   \\
           & $1^{+}$  & 0.&3   & 0.&3   & 0.&3   & 0.&3   & 0.&3   & 0.&4   & 0.&3   & 0.&3   & 0.&3   \\
\hline                                                                               
           & Total    & 227.&5 & 228.&6 & 218.&1 & 213.&8 & 227.&6 & 226.&1 & 227.&3 & 228.&0 & 248.&1 \\[2mm]
\hline                                                                               
$\Delta 2p$&          & 19.&9  & 19.&6  & 18.&0  & 17.&0  & 18.&7  & 18.&7  & 19.&2  & 18.&9  & 21.&1  \\
\hline \hline
\end{tabular}}
\end{table*}

\newpage

\subsection{2.3 $\boldsymbol{\mu}$-\ce{^{208}Pb}}

\begin{table*}[!hbtp]
\caption{Contributions from different nuclear excitation modes to the nuclear-polarization~(NP) corrections (absolute values $|\Delta E^{\mathrm{NP}}|=-\Delta E^{\mathrm{NP}}$, in~eV) to the states $1s_{1/2}$, $2p_{1/2}$ and $2p_{3/2}$ in muonic \ce{^{208}Pb}. The quantity $\Delta 2p^{\mathrm{NP}} = |\Delta E_{2p_{1/2}}^{\mathrm{NP}}| - |\Delta E_{2p_{3/2}}^{\mathrm{NP}}|$ is the corresponding NP contribution to the fine-structure splitting. The values are given for 9 different Skyrme parametrizations.}
{\renewcommand{\arraystretch}{1.35}
\renewcommand{\tabcolsep}{0.3636cm}
\begin{tabular}{lcr@{}lr@{}lr@{}lr@{}lr@{}lr@{}lr@{}lr@{}lr@{}l}
\hline \hline
           &         & \multicolumn{2}{c}{KDE0}    & \multicolumn{2}{c}{SKX}     & \multicolumn{2}{c}{SLy5}    & \multicolumn{2}{c}{BSk14}   & \multicolumn{2}{c}{SAMi}    & \multicolumn{2}{c}{NRAPR}   & \multicolumn{2}{c}{SkP}     & \multicolumn{2}{c}{SkM*}    & \multicolumn{2}{c}{SGII}    \\
\hline                                                                                                           
$1s_{1/2}$ & $0^{+}$ & 1335.&7 & 1214.&2 & 1379.&4 & 1323.&1 & 1356.&7 & \phantom{)}1437.&8 & 1473.&3 & 1462.&1 & 1465.&1 \\
           & $1^{-}$ & 2189.&4 & 2177.&4 & 2235.&8 & 2277.&5 & 2381.&5 & 2364.&7 & 2262.&7 & 2401.&3 & 2467.&1 \\
           & $2^{+}$ & 1117.&5 & 1190.&9 & 1121.&2 & 1163.&0 & 1128.&0 & 1150.&5 & 1234.&5 & 1194.&5 & 1205.&7 \\
           & $3^{-}$ & 527.&8  & 538.&3  & 529.&0  & 526.&0  & 563.&0  & 623.&2  & 534.&0  & 542.&3  & 580.&0  \\
           & $4^{+}$ & 188.&9  & 200.&1  & 188.&3  & 193.&8  & 191.&4  & 196.&9  & 200.&2  & 197.&9  & 205.&1  \\
           & $5^{-}$ & 91.&1   & 97.&2   & 91.&2   & 92.&0   & 92.&9   & 97.&8   & 96.&5   & 94.&1   & 99.&1   \\
           & $1^{+}$ & 12.&6   & 14.&0   & 12.&3   & 12.&9   & 13.&5   & 18.&1   & 13.&7   & 12.&9   & 13.&4   \\
\hline                                                                                                 
           & Total   & 5462.&9 & 5432.&0 & 5557.&2 & 5588.&2 & 5727.&0 & 5889.&0 & 5815.&0 & 5905.&1 & 6035.&5 \\[2mm]
\hline                                                                                                 
$2p_{1/2}$ & $0^{+}$ & 92.&0   & 85.&1   & 95.&8   & 91.&5   & 95.&0   & 99.&9   & 101.&5  & 101.&0  & 101.&2  \\
           & $1^{-}$ & 879.&2  & 874.&6  & 916.&2  & 966.&6  & 990.&3  & 968.&0  & 936.&8  & 1032.&6 & 1042.&8 \\
           & $2^{+}$ & 494.&3  & 559.&9  & 503.&4  & 525.&0  & 508.&3  & 543.&8  & 591.&9  & 544.&0  & 546.&9  \\
           & $3^{-}$ & 218.&7  & 226.&6  & 221.&5  & 217.&6  & 242.&5  & 279.&1  & 221.&5  & 225.&3  & 246.&2  \\
           & $4^{+}$ & 63.&7   & 68.&4   & 64.&2   & 66.&1   & 66.&1   & 68.&3   & 68.&2   & 67.&8   & 70.&3   \\
           & $5^{-}$ & 29.&2   & 31.&2   & 29.&6   & 29.&7   & 30.&6   & 32.&4   & 31.&1   & 30.&6   & 32.&3   \\
           & $1^{+}$ & 3.&6    & 4.&1    & 3.&2    & 3.&8    & 3.&7    & 5.&1    & 4.&0    & 3.&8    & 4.&0    \\
\hline                                                                                                 
           & Total   & 1780.&8 & 1850.&1 & 1833.&8 & 1900.&3 & 1936.&5 & 1996.&5 & 1955.&0 & 2005.&1 & 2043.&7 \\[2mm]
\hline                                                                                                 
$2p_{3/2}$ & $0^{+}$ & 53.&6   & 49.&8   & 56.&0   & 53.&4   & 55.&7   & 58.&5   & 59.&4   & 59.&0   & 59.&1   \\
           & $1^{-}$ & 886.&0  & 882.&4  & 922.&7  & 981.&3  & 995.&7  & 968.&7  & 932.&0  & 1037.&2 & 1050.&0 \\
           & $2^{+}$ & 493.&3  & 561.&0  & 502.&6  & 524.&2  & 508.&1  & 551.&7  & 594.&4  & 542.&8  & 545.&6  \\
           & $3^{-}$ & 205.&8  & 213.&1  & 208.&4  & 204.&8  & 228.&2  & 262.&5  & 208.&2  & 211.&8  & 231.&5  \\
           & $4^{+}$ & 58.&2   & 62.&3   & 58.&6   & 60.&3   & 60.&4   & 62.&4   & 62.&1   & 61.&8   & 64.&1   \\
           & $5^{-}$ & 26.&1   & 27.&8   & 26.&5   & 26.&5   & 27.&4   & 29.&0   & 27.&6   & 27.&3   & 28.&9   \\
           & $1^{+}$ & 1.&7    & 1.&9    & 1.&6    & 1.&7    & 2.&0    & 3.&2    & 1.&8    & 1.&7    & 1.&9    \\
\hline                                                                                                 
           & Total   & 1724.&7 & 1798.&3 & 1776.&4 & 1852.&3 & 1877.&4 & 1936.&0 & 1885.&6 & 1941.&8 & 1981.&0 \\[2mm]
\hline                                                                                                 
$\Delta 2p$&         & 56.&0   & 51.&8   & 57.&5   & 48.&1   & 59.&1   & 60.&5   & 69.&3   & 63.&3   & 62.&7   \\
\hline \hline
\end{tabular}}
\end{table*}

\newpage

\begin{table*}[!hbtp]
\caption{Contributions from different nuclear excitation modes to the nuclear-polarization~(NP) corrections (absolute values $|\Delta E^{\mathrm{NP}}|=-\Delta E^{\mathrm{NP}}$, in~eV) to the states $3p_{1/2}$ and $3p_{3/2}$ in muonic \ce{^{208}Pb}. The quantity $\Delta 3p^{\mathrm{NP}} = |\Delta E_{3p_{1/2}}^{\mathrm{NP}}| - |\Delta E_{3p_{3/2}}^{\mathrm{NP}}|$ is the corresponding NP contribution to the fine-structure splitting. The values are given for 9 different Skyrme parametrizations.}
{\renewcommand{\arraystretch}{1.35}
\renewcommand{\tabcolsep}{0.4040cm}
\begin{tabular}{lcr@{}lr@{}lr@{}lr@{}lr@{}lr@{}lr@{}lr@{}lr@{}l}
\hline \hline
           &         & \multicolumn{2}{c}{KDE0}    & \multicolumn{2}{c}{SKX}     & \multicolumn{2}{c}{SLy5}    & \multicolumn{2}{c}{BSk14}   & \multicolumn{2}{c}{SAMi}    & \multicolumn{2}{c}{NRAPR}   & \multicolumn{2}{c}{SkP}     & \multicolumn{2}{c}{SkM*}    & \multicolumn{2}{c}{SGII}    \\
\hline                                                                                                           
$3p_{1/2}$ & $0^{+}$ & 33.&7   & 31.&0   & 35.&1   & \phantom{0.}33.&5   & 34.&7   & \phantom{0).}36.&6   & 37.&3   & 37.&1   & 37.&1   \\
           & $1^{-}$ & 247.&4  & 263.&6  & 267.&8  & 273.&3  & 316.&3  & 229.&4  & 289.&6  & 307.&8  & 311.&6  \\
           & $2^{+}$ & 154.&0  & 182.&4  & 158.&2  & 164.&0  & 162.&7  & 156.&8  & 204.&0  & 171.&9  & 172.&5  \\
           & $3^{-}$ & 65.&3   & 67.&7   & 65.&9   & 64.&9   & 72.&4   & 84.&1   & 65.&9   & 67.&1   & 73.&6   \\
           & $4^{+}$ & 19.&0   & 20.&4   & 19.&1   & 19.&7   & 19.&6   & 20.&3   & 20.&3   & 20.&2   & 20.&9   \\
           & $5^{-}$ & 8.&8    & 9.&4    & 8.&9    & 8.&9    & 9.&2    & 9.&7    & 9.&3    & 9.&2    & 9.&7    \\
           & $1^{+}$ & 1.&2    & 1.&4    & 1.&0    & 1.&2    & 1.&2    & 2.&9    & 1.&3    & 1.&2    & 1.&3    \\
\hline                                                                                                 
           & Total   & 529.&3  & 575.&9  & 556.&0  & 565.&5  & 616.&2  & 539.&7  & 627.&7  & 614.&4  & 626.&7  \\[2mm]
\hline                                                                                                 
$3p_{3/2}$ & $0^{+}$ & 20.&6   & 19.&1   & 21.&6   & 20.&5   & 21.&4   & 22.&5   & 23.&0   & 22.&8   & 22.&8   \\
           & $1^{-}$ & 278.&1  & 293.&5  & 301.&9  & 309.&6  & 346.&2  & 262.&5  & 322.&9  & 338.&5  & 349.&3  \\
           & $2^{+}$ & 167.&4  & 202.&3  & 172.&5  & 178.&3  & 179.&6  & 175.&2  & 231.&1  & 187.&6  & 188.&2  \\
           & $3^{-}$ & 65.&5   & 67.&8   & 66.&1   & 65.&1   & 72.&6   & 84.&3   & 66.&0   & 67.&2   & 73.&7   \\
           & $4^{+}$ & 18.&4   & 19.&7   & 18.&5   & 19.&1   & 19.&1   & 19.&7   & 19.&6   & 19.&5   & 20.&3   \\
           & $5^{-}$ & 8.&3    & 8.&9    & 8.&4    & 8.&5    & 8.&7    & 9.&2    & 8.&8    & 8.&7    & 9.&2    \\
           & $1^{+}$ & 0.&4    & 0.&6    & 0.&4    & 0.&4    & 0.&6    & 2.&1    & 0.&5    & 0.&5    & 0.&5    \\
\hline                                                                                                 
           & Total   & 558.&8  & 611.&8  & 589.&4  & 601.&6  & 648.&1  & 575.&5  & 671.&8  & 644.&7  & 664.&0  \\[2mm]
\hline                                                                                                 
$\Delta 3p$&         & -29.&5  & -35.&9  & -33.&4  & -36.&1  & -31.&9  & -35.&8  & -44.&1  & -30.&3  & -37.&7  \\
\hline \hline
\end{tabular}}
\end{table*}